\documentclass[aps,prd,groupedaddress,nofootinbib,superscriptaddress,showkeys,showpacs,eqsecnum,amsmath,amssymb]{revtex4-2}

\usepackage{amsmath}
\usepackage{graphicx}
\usepackage{amsfonts}
\usepackage{latexsym}
\usepackage{bbold}
\usepackage{wasysym}
\usepackage{calligra}
\usepackage{float}
\usepackage{ulem}
\usepackage{inputenc}
\usepackage{xspace}
\usepackage{url}
\usepackage{epstopdf}
\usepackage{tikz}
\usepackage{amssymb}
\usepackage{enumitem}

\usepackage{amsmath}
\usepackage{amssymb}
\usepackage{graphicx}
\usepackage{amsfonts}
\usepackage{latexsym}
\usepackage{bbold}
\usepackage{wasysym}
\usepackage{calligra}
\usepackage{float}
\usepackage{ulem}
\usepackage{inputenc}
\usepackage{xspace}
\usepackage{url}
\usepackage{epstopdf}
\usepackage{tikz}
\usepackage{amsthm}
\usepackage{soul}
\usepackage{hyperref}
\usepackage[small,nohug,heads=littlevee]{diagrams}
\diagramstyle[labelstyle=\scriptstyle]

\newtheorem{corr}{Corollary}

\newcommand{\be}{\begin{equation}}
\newcommand{\ee}{\end{equation}}
\newcommand{\beq} {\begin{equation}}
\newcommand{\eeq} {\end{equation}}
\newcommand{\ba}{\begin{eqnarray}}
\newcommand{\ea}{\end{eqnarray}}

\newcommand{\RR}[2]{R#1{}^{(#2)}}
\newcommand{\Gam}[2]{\Gamma#1{}^{(#2)}}
\newcommand{\NN}[2]{N#1{}^{(#2)}}
\newcommand{\PP}[2]{P#1{}^{(#2)}}
\newcommand{\XX}[2]{\Xi#1}

\begin{document}

	\title{Biconnection gravity as a statistical manifold}

	\author{Damianos Iosifidis}
    \email{damianos.iosifidis@ut.ee}
	\affiliation{Laboratory of Theoretical Physics, Institute of Physics, University of Tartu, W. Ostwaldi 1, 50411 Tartu, Estonia.}

    \author{Konstantinos Pallikaris}
    \email{konstantinos.pallikaris@ut.ee}
	\affiliation{Laboratory of Theoretical Physics, Institute of Physics, University of Tartu, W. Ostwaldi 1, 50411 Tartu, Estonia.}

	\date{\today}
	\begin{abstract}
		
		We formulate a biconnection theory of gravity whose gravitational action consists of a recently defined mutual curvature scalar. Namely, we build a gravitational theory consisting of one metric and two affine connections, in a metric-affine gravity setup. Consequently, coupling the two connections on an equal footing with matter, we show that the geometry of the resulting theory is, quite intriguingly, that of statistical manifolds.  This ultimately indicates a remarkable mathematical correspondence between gravity  and information geometry. 
		
	\end{abstract}
	
	\maketitle
	
	\allowdisplaybreaks
	
	
	\tableofcontents

\normalem

\section{Introduction}

While General Relativity (GR) has achieved remarkable success in explaining a wide range of phenomena, it encounters limitations when confronted with, e.g., the accelerated expansion of the Universe~\cite{SupernovaSearchTeam:1998fmf}, the existence of dark matter and dark energy~\cite{Clowe:2006eq}, or the quantum nature of gravity. Therefore, the search for compelling alternatives to GR, motivated by the need to address cosmological observations and unresolved questions that challenge the current understanding of gravity, is always a justified course of action~\cite{CANTATA:2021ktz}. 

Among other promising alternatives are bimetric theories of gravity,\footnote{A teleparallel bigravity analogue has been recently developed in~\cite{blixt2023teleparallel} and a two-metrics/two-connections gravity in~\cite{Gialamas:2023aim}.} not strictly in the context of nonlinear massive gravity \cite{deRham:2010ik, deRham:2010kj,Hassan:2011hr, deRham:2011rn},\footnote{For a comprehensive review of massive gravity, see~\cite{Hinterbichler:2011tt}.} but also in the more general framework of consistent (ghost-free) theories of interacting spin-2 fields, where the customarily nondynamical reference metric of massive gravity acquires its own dynamics (see~\cite{Hassan:2011zd, Hassan:2011tf} and references therein).\footnote{Historically, such a configuration was first introduced in~\cite{Isham:1971gm} to describe a (massive spin-2) meson-graviton interaction.} Nevertheless, all of the above theories assume that the connection is the unique metric-compatible and torsion-free connection induced by the metric, the Levi-Civita connection. Consequently, the only dynamical gravitational field variables in the problem are the two metrics and nothing else. 

However, the connection ought not be the Levi-Civita one. Departing from the geometry underlying GR, we encounter geometries with torsion and nonmetricity in gravitational theories where the metric and the connection are independent field variables, and variations are performed with respect to both of them in order to obtain the field equations. This framework is called Metric-Affine  Gravity (MAG)~\cite{hehl1995metric}.\footnote{Some recent developments in MAG include~\cite{Iosifidis:2023pvz, Iosifidis:2021bad, Iosifidis:2021fnq, Iosifidis:2021tvx, vitagliano2011dynamics, sotiriou2007metric,  percacci2020new, Jimenez:2020dpn, BeltranJimenez:2019acz, aoki2019scalar, Cabral:2020fax, Yang:2021fjy, Ariwahjoedi:2021yth, bahamonde2020new, Bahamonde:2022meb, bahamonde2023new, shimada2019metric, Yang:2021fjy, kubota2021cosmological, Kubota:2020ehu, Mikura:2020qhc, Mikura:2021ldx, Boudet:2022nub}.} Interestingly enough, these extended geometries (with torsion and nonmetricity) also appear in a different branch of mathematics that collectively goes by the name Information Geometry (IG)~\cite{amari1997information}.\footnote{The literature on the subject is vast. For some recent works see \cite{Khan:2021ltg,peyghan2022musical} and references therein. Note that there are also extensions to quantum systems, what is known as Quantum Information Geometry (see \cite{lambert2023classical} for a recent review on this topic).} For instance, in statistical models and, in particular, in the study of Statistical Manifolds, one encounters a geometry that, besides the metric, is also endowed with two affine connections which are dual to each other (see~\cite{amari1987differential} for more information). If these two connections are also torsion-free, then there exists a totally symmetric, usually called the {cubic tensor}~\cite{amari1997information}, which fully describes the two connections and measures the deviation of each from the Levi-Civita connection. 

It is then interesting to examine if such a geometric arena could manifest itself as a background solving the field equations in some gravitational construction. If such a {mathematical} correspondence exists, it could be used to conjecture a deeper relation between these seemingly unrelated fields of research, some sort of IG/gravity correspondence in the---loosely speaking---fashion of gauge/gravity duality. As we shall explicitly demonstrate, such a correspondence does exist and therefore, it is a good starting point to further investigate a potential interrelation between statistical models and the microproperties of matter as encoded in the hypermomentum tensor. As we will show, there is a certain hypermomentum source directly related to the so-called cubic tensor appearing in the study of statistical manifolds. 

In this regard, we shall present here what we may call a biconnection formulation of gravity. The inclusion of two affine connections is not as arbitrary as it may seem. For instance, given a nonmetric connection,
one can define a second one (its dual) in such a way that inner products are preserved even though the manifold has nonmetricity~\cite{amari1987differential}. In addition, as was recently shown by one of us in~\cite{iosifidis2023torsion}, by suitably defining the second connection (the so-called torsion-dual connection) one can ``preserve quadrilaterals'' even if both connections are torsionful. As we shall explicitly show, both of these geometries and also a more general one, can be accommodated in the biconnection formulation we develop here.

The paper is organized as follows. In section~\ref{sec:Pre}, we communicate the basic ingredients in MAG; we introduce notation and definitions we are going to use in the rest of this work. We then formulate three distinct cases of biconnection theories and show the correspondences with the statistical structures for certain forms of hypermomentum. In particular, we start with the case of a symmetric connection, then we investigate the case with a metric connection, and we finally generalize the formulation for general unconstrained connections in a general MAG-like framework. We then wrap up our main results and point to future directions.

\section{Preliminaries \label{sec:Pre}}
In MAG, the connection $\nabla$ is an independent gravitational entity besides the metric tensor. Associated to it is a set of coefficients, denoted by $\Gamma^\lambda{}_{\mu\nu}$, and a covariant derivative $\nabla_\mu$ which acts on tensors $\mathsf{T}$ of arbitrary rank $(p,q)$ in the following way, 
\begin{eqnarray}
    \nabla_\mu T^{\alpha_1...\alpha_p}{}_{\beta_1...\beta_q} &=& \partial_\mu T^{\alpha_1...\alpha_p}{}_{\beta_1...\beta_q}+\Gamma^{\alpha_1}{}_{\nu\mu}T^{\nu\alpha_2...\alpha_p}{}_{\beta_1...\beta_q}+\Gamma^{\alpha_2}{}_{\nu\mu}T^{\alpha_1\nu...\alpha_p}{}_{\beta_1...\beta_q}+...-\nonumber\\
    &&-\Gamma^\nu{}_{\beta_1\mu}T^{\alpha_1...\alpha_p}{}_{\nu\beta_2...\beta_q}-\Gamma^\nu{}_{\beta_2\mu}T^{\alpha_1...\alpha_p}{}_{\beta_1\nu...\beta_q}-...\;.
\end{eqnarray}
Torsion and nonmetricity are intrinsic properties of the connection, given by 
\begin{equation}
    S_{\mu\nu}{}^\lambda = \Gamma^\lambda{}_{[\mu\nu]}\quad\text{and}\quad  Q_{\lambda\mu\nu}=-\nabla_\lambda g_{\mu\nu},
\end{equation}
respectively. The torsion tensor measures the failure of the connection to be symmetric in its lower indices, and it is associated with the inability to ``close quadrilaterals'' via parallel transport of vectors along curves in spacetime. The nonmetricity tensor measures the failure of the metric to be covariantly constant; the norms of vectors (or more generally, their inner products) change as they are transported along a path. By contracting torsion and nonmetricity with the Kronecker delta and the metric, we obtain three vectors, namely
\begin{equation}
    S_\mu=S_{\mu\nu}{}^\nu,\quad Q_\mu = Q_{\mu\nu\lambda}g^{\nu\lambda},\quad \text{and}\quad q_{\mu} = Q_{\nu\mu\lambda}g^{\lambda\nu}.
\end{equation}
The middle vector in the above often goes by the name ``Weyl vector'' in the literature. 

Continuing, the curvature tensor associated with $\nabla$ is given by 
\begin{equation}
    R^\lambda{}_{\rho\mu\nu} = 2\partial_{[\mu}\Gamma^\lambda{}_{|\rho|\nu]}+2\Gamma^\lambda{}_{\sigma[\mu}\Gamma^{\sigma}{}_{|\rho|\nu]}.
\end{equation}
It admits three distinct single traces, the Ricci tensor $R_{\mu\nu}=R^\lambda{}_{\mu\lambda\nu}$, the co-Ricci tensor $\check{R}^\mu{}_{\nu} = R^\mu{}_{\lambda\nu\rho}g^{\lambda\rho}$, and the homothetic (curvature) tensor $\hat{R}_{\mu\nu}=R^\lambda{}_{\lambda\mu\nu}=\partial_{[\mu}Q_{\nu]}$. It also has a unique double trace, $R=R_{\mu\nu}g^{\mu\nu}=\check{R}^\mu{}_\mu$ which is known as the Ricci scalar. Another important quantity in metric-affine theories is the distortion tensor
\begin{equation}
    N^\lambda{}_{\mu\nu} = \Gamma^\lambda{}_{\mu\nu} - \tilde{\Gamma}^\lambda{}_{\mu\nu},\label{eq:DistortionTensor}
\end{equation}
which measures the difference between the full connection and the Levi-Civita connection $\tilde{\nabla}$, the latter associated with a set of coefficients 
\begin{equation}
    \tilde{\Gamma}^\lambda{}_{\mu\nu} = \frac12 g^{\lambda\rho}\left( \partial_\mu g_{\rho\nu}+\partial_\nu g_{\rho\mu}-\partial_\rho g_{\mu\nu} \right).
\end{equation}
In terms of torsion and nonmetricity, we have that 
\begin{equation}
    N^\lambda{}_{\mu\nu} = \frac{1}{2} g^{\lambda\rho}\left(Q_{\mu\nu\rho} + Q_{\nu\rho\mu}
- Q_{\rho\mu\nu}\right)+g^{\lambda\rho}\left( S_{\mu\nu\rho} - S_{\rho\mu\nu} - S_{\nu\mu\rho}\right).\label{eq:DistortionExpansion}
\end{equation}

Rearranging Eq.~\eqref{eq:DistortionTensor}, we see that we can always express the general connection in terms of the Levi-Civita connection and the distortion tensor. Making this trade results in the decomposition of connection-dependent quantities into Riemannian and non-Riemannian parts, what is known as post-Riemannian expansion. For example, the post-Riemannian expansion of the curvature tensor reads 
\begin{equation}
R^\lambda{}_{\rho \mu\nu}  =  \tilde{R}^\lambda{}_{ \rho \mu\nu} +  2\tilde{\nabla}_{[\mu} {N^\lambda}_{|\rho|\nu]} + 2{N^\lambda}_{\sigma[\mu} {N^\sigma}_{|\rho|\nu]} ,
\end{equation}
where $\tilde{\nabla}_\mu$ is the covariant derivative associated with the Levi-Civita connection, and $\tilde{R}^\lambda{}_{\rho\mu\nu}$ is the familiar Riemann tensor (the curvature tensor of the Levi-Civita connection).\footnote{In this manuscript, unless otherwise stated, quantities with a tilde accent will always stand for objects associated with the Levi-Civita connection.}

Now, if we equip our manifold with two connections, say $\{\nabla^{(i)}\}_{i=1,2}$, associated with two sets of coefficients $\{ \Gam{^\lambda{}_{\mu\nu}}{i}\}$, their difference is a true tensor 
\begin{equation}
    K^\lambda{}_{\mu\nu}:= \Gam{^\lambda{}_{\mu\nu}}{1} - \Gam{^\lambda{}_{\mu\nu}}{2},\label{eq:diften}
\end{equation}
called the {difference tensor}. Note that the previously mentioned distortion tensor is a particular case of the latter tensor for $\Gamma^\lambda{}_{\mu\nu}{}^{(2)}\equiv \tilde{\Gamma}^\lambda{}_{\mu\nu}$.
In addition, employing a post-Riemannian expansion for both connections, the Levi-Civita terms drop out and we have the equivalent expression
\beq
  K^\lambda{}_{\mu\nu}=\NN{^{\lambda}{}_{\mu\nu}}{1}-\NN{^{\lambda }{}_{\mu\nu}}{2}
\eeq
Interestingly, the difference tensor $K^\lambda{}_{\mu\nu}$ does naturally appear when acting with the commutator of the two covariant derivatives upon a scalar function $f$, 
\begin{equation}
   [\nabla^{(1)}_{\mu},\nabla^{(2)}_{\nu}]f=-K^\lambda{}_{\nu\mu}\partial_\lambda f.
\end{equation}
Notice that the sum of two connections does not define a new connection. However, the convex linear combination
\begin{equation}
    \nabla = t \nabla^{(1)}+(1-t)\nabla^{(2)},\label{eq:conaddition}
\end{equation}
with $t$ being a real parameter, does transform as a connection. Its curvature is given by 
\begin{equation}
    R^\lambda{}_{\rho\mu\nu} = t \RR{^{\lambda}{}_{\rho\mu\nu}}{1} + (1-t)\RR{^\lambda{}_{\rho\mu\nu}}{2}-2t(1-t)K^\lambda{}_{\sigma[\mu}K^\sigma{}_{|\rho|\nu]}.\label{eq:CurvatureConvex}
\end{equation}
Nevertheless, a more interesting object is the so-called {mutual curvature} of the two connections for which a self-consistent definition was given in~\cite{iosifidis2023torsion}. In local coordinates, it reads
\beq
\mathcal{R}^{\lambda}{}_{\rho\mu\nu}=\frac{1}{2}\left(\RR{^\lambda{}_{\rho\mu\nu}}{1}+\RR{^\lambda{}_{\rho\mu\nu}}{2}\right)-K^\lambda{}_{\sigma[\mu}K^{\sigma}{}_{|\rho|\nu]}.\label{eq:mutualcurvature}
\eeq
The geometric meaning of the mutual curvature can be found in Appendix \ref{sec:PT}. The above definition will be the basic ingredient for our biconnection theory. Observe that $\mathcal{R}^\lambda{}_{\rho\mu\nu}$ is not a curvature tensor as per the standard definition since it is not associated with a connection; there is no value of $t$ such that Eq.~\eqref{eq:CurvatureConvex} assumes the form~\eqref{eq:mutualcurvature}.

In fact, this last definition possesses a notable advantage, in comparison to (\ref{eq:CurvatureConvex}). The latter is invariant under the simultaneous changes $\Gam{^\lambda{}_{\mu\nu}}{1}\mapsto \Gam{^\lambda{}_{\mu\nu}}{1}+(1-t)\Omega^{\lambda}{}_{\mu\nu}$ and $\Gam{^\lambda{}_{\mu\nu}}{2} \mapsto \Gam{^\lambda{}_{\mu\nu}}{2}-t\Omega^{\lambda}{}_{\mu\nu}$. 
This last invariance suggests that a biconnection theory, constructed out of (\ref{eq:CurvatureConvex}), would be equivalent (up to gauge) to a theory for a single independent connection, since one of the two distortion tensors would be pure gauge. In other words, one of the connections would always be the Levi-Civita connection (in a certain gauge), meaning that the difference tensor would just be the usual distortion tensor. Having formulated the appropriate setup let us proceed with the model building of Metric-Affine biconnection Theories.

\section{biconnection theory}

In the following we shall formulate three different biconnection Theories and also show their correspondences (under certain conditions) with statistical and torsion-statistical Manifolds. In order to cover all possibilities we shall first formulate a symmetric (i.e. vanishing torsion)  theory, followed by a metric one (i.e. vanishing nonmetricity) and finally, generalizing, we shall construct a general metric-affine biconnection theory.

\subsection{Gravity as a statistical manifold}

We consider the gravitational part of our biconnection theory to be given by
 \beq
S[g,\Gam{}{1},\Gam{}{2}]=\frac{1}{2\kappa}\int d^{n}x \sqrt{-g}\mathcal{R} \label{eq:BCactionGrav}
 \eeq
where $\mathcal{R}=\mathcal{R}^\lambda{}_{\mu\lambda\nu}g^{\mu\nu}$ and $\kappa=8\pi G_n$ with $G_n$ being the gravitational constant in $n$ dimensions. The two connections are assumed to be torsion-free. Using relation (\ref{eq:mutualcurvature}) and taking traces in order to form the scalar curvatures, we may equivalently write~\eqref{eq:BCactionGrav} as 
\beq
S=\frac{1}{4\kappa}\int d^{n}x \sqrt{-g}\left(\RR{}{1}+\RR{}{2}+K\right),
\eeq
where $\{\RR{}{i}\}_{i=1,2}$ are the scalar curvatures of the two connections formed by the curvatures associated with $\nabla^{(1)}$ and $\nabla^{(2)}$, respectively, and
 \beq
K:=\left( K^{\lambda\mu\nu}K_{\mu\nu\lambda}-K^{\lambda\mu}{}_{\mu}K^{\nu}{}_{\lambda \nu}\right)
 \eeq
is what we shall call the {difference scalar}. Note that in our approach we do not simply consider two affine connections, compute the associated Ricci scalars for each one, and merely add them up. This procedure which could be called a decoupled biconnection gravity was considered in \cite{Khosravi:2013kha}.  Here, we rather start with the mutual curvature scalar as was defined in~\cite{iosifidis2023torsion}. This leads to a quite interesting result since, apart from the Ricci scalars of the two connections which do not interact, we have couplings between the two connections via the terms quadratic in the difference tensor $K^{\lambda}_{\;\;\mu\nu}$.\footnote{The situation is analogous to the total magnetic energy formula of two current-driven circuits in Electromagnetism. Recall there that, since the two systems interact, the energy does not have the additive property. Besides the individual magnetic energies, there is also a mixing term appearing which is proportional to the the mutual inductance multiplied by the two currents. One could then roughly say, by crude analogy, that the mutual-inductance term in the aforementioned configuration corresponds to the difference scalar in our setup.} 

Let us note here that one could just as well formulate the biconnection theory as a single connection theory with a metric and an additional rank-3 tensor field. Indeed, this is always possible and a similar situation appears also in the usual metric-affine formulation. There, instead of the metric and an independent connection one can trade variables and use the metric, the Levi-Civita connection induced by it, and an additional rank-3 tensor field which is the distortion tensor (see~\cite{Iosifidis:2023pvz} for details). In addition, here one may alternatively formulate the biconnection theory either as a field theory for a single connection, the metric and a rank-3 tensor field, as mentioned above, or as a field theory for a metric and two additional rank-3 tensor fields (the distortions of the two connections); matter will then couple to their difference.

For the variations of the difference scalar with respect to $\Gamma^{(1)}$, $\Gamma^{(2)}$, we readily find
\begin{subequations}
\begin{eqnarray}
\delta_{\Gamma^{(1)}}K&=&\delta \Gam{^{\lambda}{}_{\mu\nu}}{1}\Psi_{\lambda}{}^{(\mu\nu)},\\
\delta_{\Gamma^{(2)}}K&=&-\delta \Gam{^{\lambda}{}_{\mu\nu}}{2}\Psi_{\lambda}{}^{(\mu\nu)},
\end{eqnarray}
\end{subequations}
respectively, where 
\begin{equation}
    \Psi_\lambda{}^{\mu\nu}:=K^{\nu}{}_\lambda{}^{\mu}+K^{\mu\nu}{}_{\lambda}-K^{\alpha}{}_{\lambda \alpha}g^{\mu\nu}-K^{\mu\alpha}{}_{\alpha}\delta_{\lambda}^{\nu}.\label{psi}
\end{equation}
With these we can then find the connection field equations. Varying the action~\eqref{eq:BCactionGrav} with respect to the two affine connections (see Appendix~\ref{AppA} for some useful formulas), we obtain the connection field equations
\begin{subequations}\label{eq:ConnFE}
\begin{eqnarray}
\PP{_\lambda{}^{\mu\nu}}{1}+\Psi_{\lambda}{}^{(\mu\nu)}&=&0,\\ 
\PP{_\lambda{}^{\mu\nu}}{2}-\Psi_{\lambda}{}^{(\mu\nu)}&=&0,
\end{eqnarray}
\end{subequations}
where 
\begin{eqnarray}
P_{\lambda}{}^{\mu\nu(i)}&=& \frac{1}{2}Q_{\lambda}{}^{(i)}g^{\mu\nu}-Q_{\lambda}{}^{\mu\nu(i)}+\delta_{\lambda}^{(\nu} \left(q^{\mu)(i)}-\frac{1}{2}\,Q^{\mu) (i)}\right)\label{PalatiniQ}
\end{eqnarray}
is the torsion-free Palatini tensor associated with the $i$-th connection. 

Adding the two, it follows that
\beq
P_{\lambda}^{\;\;\mu\nu(1)}+P_{\lambda}^{\;\;\mu\nu(2)}=0
\eeq
After removing traces, this equation tells us that 
\beq
\nabla_{\alpha}g_{\mu\nu}=\frac{1}{2}\Big(\nabla_{\alpha}{}^{(1)}+\nabla_{\alpha}{}^{(2)}\Big)g_{\mu\nu}=0
\eeq
Therefore, the connection $\nabla$, that is the mean connection, is compatible with the metric. As noted in the introduction, using the decompositions
\begin{equation}
    \Gam{^\lambda{}_{\mu\nu}}{i}=\tilde{\Gamma}^\lambda{}_{\mu\nu}+\NN{^\lambda{}_{\mu\nu}}{i},\label{eq:ConnDeco}
\end{equation}
it immediately follows that the difference tensor takes the form
\beq
K^{\alpha\mu\nu}=N^{\alpha\mu\nu(1)}-N^{\alpha\mu\nu(2)},
\eeq 
which, when substituted  into (\ref{psi}), results in the useful relation
\beq
\Psi_{\lambda}{}^{(\mu\nu)}=P_{\lambda}{}^{\mu\nu(2)}-P_{\lambda}{}^{\mu\nu(1)} \label{psiNN}.
\eeq
Using this, the field equations~\eqref{eq:ConnFE} can be written as
\beq
P^{\alpha\mu\nu(2)}=0\quad \text{and} \quad P^{\alpha\mu\nu(1)}=0,\label{eq:ConnFEalt}
\eeq
respectively. 

It is not difficult to show that the above constraints imply that both connections are metric-compatible and since they are also torsion-free by assumption, these will coincide with the Levi-Civita connection. To see this we first take the two distinct traces\footnote{Note that $P^{\alpha\mu\nu(i)}$ is symmetric in the last two indices since we started with vanishing torsion tensors. Therefore, only two traces of the latter are truly independent.} of the above two equations which, when combined, imply that
\beq
Q_{\mu}{}^{(i)}=0=q_{\mu}{}^{(i)}.
\eeq
Further substitution of these into (\ref{eq:ConnFEalt})yields
\beq
Q_{\alpha\mu\nu}{}^{(i)}=0,
\eeq
exactly as stated above.

Consequently, the biconnection theory in vacuum is indistinguishable from GR, for the metric field equations take the usual form
\beq
\tilde{R}_{\mu\nu}-\frac{\tilde{R}}{2}g_{\mu\nu}=0,
\eeq
where $\tilde{R}_{\mu\nu}$ and $\tilde{R}$ are the Riemannian (i.e. computed with respect to the Levi-Civita connection) Ricci tensor and scalar, respectively. Not surprisingly, in order to get deviations from GR one would have to include connection-matter couplings which after all constitute an essential feature of the  metric-affine framework, relating microscopic characteristics of matter to the generalized geometry. This is what we consider in the following subsection. 

\subsubsection{Adding connection-matter couplings}

Since we are considering the framework where the two affine connections and the metric are totally independent, in the presence of matter, it is quite essential to consider connection-matter couplings. In the metric-affine formulation such couplings are very important, bringing about the so-called hypermomentum tensor~\cite{hehl1976hypermomentum} which describes the microproperties of matter as noted many times by now. 

Therefore, we consider matter which not only couples to the metric, but also to the  affine connection. In our biconnection formulation it is then logical to ask which connection couples to matter and in what way. As we have probably made it clear already, in our formulation we would like to treat both connections on equal footing since we started with the mutual scalar curvature as our Gravitational action. Therefore, we will couple matter to both connections. But how can we do it in such a way that we do not discriminate between the two? The answer is hidden in the difference tensor. Indeed, the difference tensor, as given by Eq. (\ref{eq:diften}), places the two connections on equal footing, for it is symmetric under the exchange $\{\Gamma^{(1)},\Gamma^{(2)}\}\mapsto - \{\Gamma^{(2)},\Gamma^{(1)}\}$. Consequently, it is natural to assume that the connection-matter couplings appear only as matter couplings to the difference tensor. 

In other words, the matter sector of the biconnection theory will read 
\beq
S_{\mathrm{M}}[g,\Gamma^{1},\Gamma^{2},\phi]=S_{\mathrm{M}}[g,K,\phi]=\int d^{n}x \sqrt{-g}\mathcal{L}_{\mathrm{M}}(g,K,\phi), \label{coup}
\eeq
where $\phi$ collectively denotes arbitrarily many matter fields.\footnote{The argument $K$ of $S_{\mathrm{M}}$ refers to the difference {tensor}, and should not be confused with the difference scalar appearing above.} The two hypermomenta associated with the above action are easily computed to be
\beq
\Delta_{\lambda}^{\;\;\mu\nu(1)}:=-\frac{2}{\sqrt{-g}}\frac{\delta S_{\mathrm{M}}}{\delta \Gamma^{\lambda}_{\;\;\mu\nu}{}^{(1)}}=\Xi_{\lambda}^{\;\;\mu\nu}
\eeq
and
\beq
\Delta_{\lambda}^{\;\;\mu\nu(2)}:=-\frac{2}{\sqrt{-g}}\frac{\delta S_{\mathrm{M}}}{\delta \Gamma^{\lambda}_{\;\;\mu\nu}{}^{(2)}}=-\Xi_{\lambda}^{\;\;\mu\nu},
\eeq
respectively, where we have used equation (\ref{eq:diften}), and we also invoked the chain rule. In addition, we have  defined the ``principle''  hypermomentum 
\beq
\Xi_{\lambda}^{\;\;\mu\nu}:=-\frac{2}{\sqrt{-g}}\frac{\delta S_{\mathrm{M}}}{\delta K^{\lambda }_{\;\;\mu\nu}},\label{eq:XiDef}
\eeq
which is the one ultimately appearing in the connection field equations. Bear in mind that it is symmetric in the last two indices due to vanishing torsions. Therefore, our full action reads
\beq
S[g,\Gamma^{(1)},\Gamma^{(2)},\phi]=\int d^{n}x \sqrt{-g}\Big[ \frac{1}{2\kappa}\mathcal{R}+\mathcal{L}_{\mathrm{M}}(g,K,\phi) \Big] \label{act}
\eeq

Varying with respect to the connections, and using the above definitions, we easily obtain the field equations
\begin{subequations}
\begin{eqnarray}
P_{\lambda}^{\;\;\mu\nu(1)}+\Psi_{\lambda}^{\;\;(\mu\nu)}&=&2\kappa\Xi_{\lambda}^{\;\;\mu\nu},\\
P_{\lambda}^{\;\;\mu\nu(2)}-\Psi_{\lambda}^{\;\;(\mu\nu)}&=&-2\kappa\Xi_{\lambda}^{\;\;\mu\nu}.
\end{eqnarray}
\end{subequations}
Using Eq. (\ref{psiNN}), the above are written as
\beq
P_{\lambda}^{\;\; \mu\nu(2)}=2\kappa \Xi_{\lambda}^{\;\;\mu\nu}
\eeq
and 
\beq
P_{\lambda}^{\;\; \mu\nu(1)}=-2\kappa \Xi_{\lambda}^{\;\;\mu\nu}.
\eeq
Using the definition~\eqref{PalatiniQ} of the torsion-free $P_{\lambda}^{\;\;\mu\nu(i)}$, we may easily find the general solutions (nonmetricities in terms of hypermomentum) to the above, which read 
\beq
Q_{\alpha\mu\nu}{}^{(2)}=-2\kappa\Xi_{\alpha\mu\nu}+\frac{2\kappa}{n-2}\left( \Xi_{\alpha\beta}{}^\beta-\frac{2}{n-1}\Xi^\beta{}_{\alpha\beta}\right) g_{\mu\nu}+\frac{4 \kappa}{(n-1)}\Xi_{\beta(\mu}{}^\beta g_{\nu)\alpha}
\eeq
and 
\beq
Q_{\alpha\mu\nu}{}^{(1)}=2\kappa\Xi_{\alpha\mu\nu}-\frac{2\kappa}{n-2}\left( \Xi_{\alpha\beta}{}^\beta-\frac{2}{n-1}\Xi^\beta{}_{\alpha\beta}\right) g_{\mu\nu}-\frac{4 \kappa}{(n-1)}\Xi_{\beta(\mu}{}^\beta g_{\nu)\alpha},
\eeq
respectively. Note that the two nonmetricities annihilate each other, namely
\beq
Q_{\alpha\mu\nu}{}^{(1)}+Q_{\alpha\mu\nu}{}^{(2)}=0,
\eeq
exactly like in the case of the dual connections in statistical manifolds. The latter property also implies that the mean connection,
\beq
\nabla\equiv \frac{1}{2}(\nabla^{(1)}+\nabla^{(2)})=\tilde{\nabla},
\eeq
is the Levi-Civita connection. 

We will now show that for specific hypermomentum sources, biconnection gravity is in a one-to-one correspondence with the statistical-manifold structure of IG. To be more precise, the underlying geometry of this spacetime with two connections can be identified with the geometry of a (Lorentzian) statistical manifold.

\subsubsection{Gravity/statistical manifold correspondence}

Let us now assume that the hypermomentum tensor $\Xi_{\alpha\mu\nu}$ is totally symmetric and traceless\footnote{This assumption does have a physical motivation. It is known for instance that a totally symmetric and traceless nonmetricity can describe pure spin-3 particle states~\cite{baekler2006linear}.}, that is
\beq
\Xi_{\alpha\mu\nu}=\Xi_{(\alpha\mu\nu)} \;\;, \;\; \Xi^{\lambda \mu}_{\;\;\;\mu}=0=\Xi^{\mu\lambda}_{\;\;\;\mu}=\Xi_{\mu}^{\;\;\mu\lambda}\;.
\eeq
Under such circumstances, the nonmetricities of $\nabla^{(1)}$ and $\nabla^{(2)}$ read
\beq
Q_{\alpha\mu\nu}{}^{(1)}\equiv Q_{(\alpha\mu\nu)}{}^{(1)}=2\kappa \Xi_{\alpha\mu\nu}
\eeq
and 
\beq
Q_{\alpha\mu\nu}{}^{(2)}\equiv Q_{(\alpha\mu\nu)}{}^{(2)}=-2\kappa \Xi_{\alpha\mu\nu},
\eeq
respectively. These imply that the connection coefficients of the two connections are written as
\begin{subequations}
\begin{eqnarray}
\Gamma^\lambda{}_{\mu\nu}{}^{(1)}&=&\Tilde{\Gamma}^\lambda{}_{\mu\nu}+\kappa\Xi^\lambda{}_{\mu\nu},\label{con1}\\
\Gamma^\lambda{}_{\mu\nu}{}^{(2)}&=&\Tilde{\Gamma}^\lambda{}_{\mu\nu}-\kappa\Xi^\lambda{}_{\mu\nu}.\label{con2}
\end{eqnarray}
\end{subequations}
Quite remarkably then, it follows that the totally symmetric trace-free hypermomentum tensor $\Xi_{\lambda\mu\nu}$ that we have here, corresponds exactly to the so-called cubic tensor $C_{\lambda\mu\nu}$ appearing in statistical manifolds. Consequently, we are led to the following statement.

\begin{corr}
For a totally symmetric and trace-free hypermomentum tensor, the biconnection gravitational theory with action (\ref{act}) is in a one-to-one correspondence with a statistical-manifold structure. In particular, the geometry of the gravitational theory is identical to that of a statistical manifold where the role of the cubic tensor is played by the hypermomentum.
\end{corr}

The above  exceptional correspondence implies that, in principle, one could get information about gravitational phenomena by studying statistical models, and vice versa. Given the fact that the hypermomentum tensor describes the microproperties of matter, it will then be possible to get information about matter's microstructure by studying the corresponding statistical manifold.

To complete the study of the torsion-free biconnection theory, we also vary (\ref{act})  with respect to the metric to obtain the metric field equations
\beq
\mathcal{R}_{(\mu\nu)}-\frac{1}{2}g_{\mu\nu}{\mathcal{R}}=\kappa T_{\mu\nu}
\eeq
where $\mathcal{R}_{\mu\nu}=\mathcal{R}^{\lambda}{}_{\mu\lambda\nu}$ is the Ricci tensor constructed out of the mutual curvature (\ref{eq:mutualcurvature}), and 
\begin{equation}
    T_{\mu\nu}=-\frac{2}{\sqrt{-g}}\frac{\delta S_{\mathrm{M}}}{\delta g^{\mu\nu}}.
\end{equation}
Now, invoking the definition of the mutual curvature, and employing the post-Riemmanian expansions of the two connections (also using expressions (\ref{con1}) and (\ref{con2})), we finally find, after some straightforward algebra, 
\beq
\tilde{R}_{\mu\nu}-\frac{1}{2}g_{\mu\nu}\tilde{R}=\kappa T_{\mu\nu}-\kappa^{2}\left(\Xi^{\alpha\beta}_{\;\;\;\;\mu}\Xi_{\alpha\beta\nu}-\frac{1}{2}\Xi^{\alpha\beta\gamma}\Xi_{\alpha\beta\gamma}g_{\mu\nu}\right),
\eeq
which are Einstein's field equations with modified sources containing $\kappa^2$ contributions from hypermomentum, namely the microstructure of matter. The Ricci form of the above reads 
\begin{equation}
    \tilde{R}_{\mu\nu} = \kappa \left( T_{\mu\nu}-\frac{1}{n-2}g_{\mu\nu} T\right)-\kappa^2 \Xi_{\mu\alpha\beta}\Xi_{\nu}{}^{\alpha\beta}.
\end{equation}

\subsection{Gravity with fermions as a torsional statistical manifold}

In \cite{iosifidis2023torsion} one of the authors formulated the concept of a ``torsion dual connection" and, related to this, the geometry of a Torsional Statistical Manifold (TSM). Given an affine connection $\nabla$ with coefficients $\Gamma^{\lambda}_{\;\;\mu\nu}$, its torsion dual $\nabla^{*}$ is defined as the connection which cooperates with $\nabla$ in keeping infinitesimal parallelograms ``unbroken'' even though both connections are endowed with torsion. The coefficients of the torsion dual connection are given by
 \beq
 \Gamma^{\lambda \star}_{\;\;\mu\nu}=\Gamma^{\lambda }_{\;\;\nu\mu}
 \eeq
 As shown in \cite{iosifidis2023torsion}, if the two connections are metric and, in addition, their distortion tensors are antisymmetric in their last two indices, then there exists a 3-form field $A_{\lambda\mu\nu}$ such that
\begin{subequations}
\begin{eqnarray}
\Gamma_{\lambda\mu\nu}&=&\tilde{\Gamma}_{\lambda\mu\nu}+A_{\lambda\mu\nu}, \label{A1}\\
\Gamma_{\lambda\mu\nu}^{*}&=&\tilde{\Gamma}_{\lambda\mu\nu}-A_{\lambda\mu\nu} \label{A2},
\end{eqnarray}
\end{subequations}
where $\tilde{\Gamma}_{\lambda\mu\nu}$ is the Levi-Civita connection and $A_{\lambda\mu\nu}=A_{[\lambda\mu\nu]}$. We shall show below how such a geometry appears naturally when one couples fermions to our biconnection formulation. 

Again, we consider the mutual scalar curvature, formed by contractions of the  mutual curvature, as the gravitational part of the action for the (now, metric) connections and allow for matter-connection couplings of the form (\ref{coup}). Then, the full action of the metric biconnection theory reads
 \beq
S[g,\Gamma^{(1)},\Gamma^{(2)},\phi]=\int d^{n}x \sqrt{-g}\Big[ \frac{1}{2\kappa}\mathcal{R}+\mathcal{L}_{\mathrm{M}}(g,K,\phi) \Big] \label{act2},
\eeq
where, let us note one more time, in contrast to (\ref{act}), where the connections were symmetric but with nonmetricity, here the connections are metric but torsionful. Varying with respect to the two connections, we obtain the field equations
\begin{subequations}
\begin{eqnarray}
\Pi_{\lambda}^{\;\;\mu\nu(1)}+\Psi_{\lambda}^{\;\;\mu\nu}&=&2\kappa\Xi_{\lambda}^{\;\;\mu\nu},\\
\Pi_{\lambda}^{\;\;\mu\nu(2)}-\Psi_{\lambda}^{\;\;\mu\nu}&=&-2\kappa\Xi_{\lambda}^{\;\;\mu\nu},
\end{eqnarray}
\end{subequations}
where
\beq
\Pi^{\alpha\mu\nu(i)}=4g^{\nu[\mu}S^{\alpha](i)}-2 S^{\alpha\mu\nu(i)}, \quad \Pi_{(\lambda\mu)\nu}=0=\Psi_{(\lambda\mu)\nu},
\eeq
with $\Psi_\lambda{}^{\mu\nu}$ defined in Eq.~\eqref{psi}.

Using the identity
\begin{equation}
    \Psi_{\lambda}{}^{\mu\nu}=\Pi_{\lambda}{}^{\mu\nu(2)}-\Pi_{\lambda}{}^{\mu\nu(1)}
\end{equation}
the above equations are written as
\beq
\Pi_{\lambda}^{\;\; \mu\nu(2)}=2\kappa \Xi_{\lambda}^{\;\;\mu\nu}
\eeq
and 
\beq
\Pi_{\lambda}^{\;\; \mu\nu(1)}=-2\kappa \Xi_{\lambda}^{\;\;\mu\nu},
\eeq
respectively, which we may combine into the single expression
\beq
\Pi_{\lambda}^{\;\; \mu\nu(i)}=(-1)^{i}2\kappa \Xi_{\lambda}^{\;\;\mu\nu}, \quad i=1,2 \label{Pi}
\eeq
We now wish to focus on connection couplings with fermions. 

For  fermionic matter, it is known (see for instance \cite{Hehl:1976kj})  that the associated hypermomentum tensor is totally antisymmetric, a 3-form field that is. In our case this translates to
\beq
\Xi_{\alpha\mu\nu}=\Xi_{[\alpha\mu\nu]}.
\eeq
Then, contraction of (\ref{Pi})  with $g_{\mu\nu}$ implies that
\beq
S_{\lambda}{}^{(1)}=0=S_{\lambda}{}^{(2)}
\eeq
which when substituted back into the same equations results in (bringing all indices up)
\beq
S^{\alpha\mu\nu(i)}={(-1)^{i+1}}\kappa \Xi^{\alpha\mu\nu}
\eeq
With this, using the expression~\eqref{eq:DistortionExpansion} for each distortion tensor (recalling also that we have vanishing nonmetricities), we find
\beq
N^{\alpha\mu\nu(i)}=(-1)^{i+1}\kappa \Xi^{\alpha\mu\nu}
\eeq
Finally, applying the decomposition rules~\eqref{eq:ConnDeco}, we find the forms of the two connections:
\begin{subequations}
\begin{eqnarray}
\Gamma_{\alpha\beta\gamma}{}^{(1)}&=&\tilde{\Gamma}_{\alpha\beta\gamma}+{\kappa}\Xi_{\alpha\beta\gamma} \label{Con1},\\
\Gamma_{\alpha\beta\gamma}{}^{(2)}&=&\tilde{\Gamma}_{\alpha\beta\gamma}-{\kappa}\Xi_{\alpha\beta\gamma} \label{Con2}.
\end{eqnarray}
\end{subequations}
         
Quite intriguingly, we see then that the role of the 3-form $A_{\alpha\mu\nu}$ is now played by the hypermomentum $\Xi_{\alpha\mu\nu}$. Subsequently, the mean connection $\nabla\equiv\frac{1}{2}(\nabla^{(1)}+\nabla^{(2)})$ is the Levi-Civita connection $\tilde{\nabla}$, and the underlying geometry is that of a TSM! Collecting everything, we arrive at: 

\begin{corr}
For  fermionic connection-matter couplings, the biconnection theory with action (\ref{act2}) is in a one-to-one correspondence with a TSM structure. In particular, the underlying geometry of the gravitational theory is identical to that of a TSM where the role of the the 3-form field $A_{\lambda\mu\nu}$ is played by the totally antisymmetric hypermomentum corresponding to fermions.
\end{corr}

To conclude the discussion on the metric biconnection theory, we also vary (\ref{act2})  with respect to the metric to again obtain the metric field equations
\beq
\mathcal{R}_{(\mu\nu)}-\frac{1}{2}g_{\mu\nu}{\mathcal{R}}=\kappa T_{\mu\nu}.
\eeq
Working as in the previous case, after some straightforward algebra, we find
\beq
\tilde{R}_{\mu\nu}-\frac{1}{2}g_{\mu\nu}\tilde{R}=\kappa T_{\mu\nu}-\kappa^{2}\left(\Xi^{\alpha\beta}_{\;\;\;\;\mu}\Xi_{\alpha\beta\nu}-\frac{1}{2}\Xi^{\alpha\beta\gamma}\Xi_{\alpha\beta\gamma}g_{\mu\nu}\right),
\eeq
which are Einstein's field equations with modified sources containing $\kappa^2$ contributions from the completely antisymmetric hypermomentum.

\subsection{General biconnection MAG }

In the previous sections, we independently considered a symmetric (i.e., vanishing torsion) and a metric (i.e., vanishing nonmetricity) biconnection theory. Hence, the connections were restricted to be torsion-free in the former case and metric-compatible in the latter.  For completeness, let us now formulate a general metric-affine biconnection theory. We shall start off with two completely general affine connections having both torsion and nonmetricity. Out of the many actions we may consider, we pick the following simple one:
\begin{equation}
    S[g,\Gam{}{1},\Gam{}{2},\phi] = \int d^nx \sqrt{-g} \left(\frac{1}{2\kappa}\mathcal{R}+\mathcal{L}_{\mathrm{M}}(g,K,\phi)\right).\label{eq:act3}
\end{equation}
This reduces to the Einstein-Hilbert action when the distortion degrees of freedom are switched off. The most general connection transformations leaving the gravitational part of the action~\eqref{eq:act3} invariant, are the projective transformations
\begin{equation}
    \Gam{^\lambda{}_{\mu\nu}}{i}\mapsto \Gam{^\lambda{}_{\mu\nu}}{i} + \delta^\lambda_\mu \xi_\nu{}^{(i)}.\label{eq:ProjTFs}
\end{equation}
which ultimately demand a vanishing dilation part $\Xi_\lambda{}^{\lambda\mu} =0$ for the hypermomentum as in the  usual (i.e. single connection) MAG formulation.

Variation of the action with respect to the connections yields the connection field equations
\begin{eqnarray}
    \PP{_\lambda{}^{\mu\nu}}{1}+\Psi_\lambda{}^{\mu\nu}=2\kappa \Xi_\lambda{}^{\mu\nu} \\
    \PP{_\lambda{}^{\mu\nu}}{2}-\Psi_\lambda{}^{\mu\nu}=-2\kappa \Xi_\lambda{}^{\mu\nu} ,\label{eq:ConnectionFE}
\end{eqnarray}
where
\begin{eqnarray}
    \PP{_{\lambda}{}^{\mu\nu}}{i} &=& -\frac{\nabla_\lambda{}^{(i)}\left(\sqrt{-g}g^{\mu\nu}\right)}{\sqrt{-g}}+\frac{\nabla_\sigma{}^{(i)}\left(\sqrt{-g}g^{\mu\sigma}\right)\delta^\nu_\lambda}{\sqrt{-g}}+2\left(S_\lambda{}^{(i)} g^{\mu\nu}-S^{\mu(i)}\delta^\nu_\lambda+g^{\mu\sigma} S_{\sigma\lambda}{}^{\nu(i)}\right),
\end{eqnarray}
and again,
\beq
\Xi_{\lambda}^{\;\;\mu\nu}:=-\frac{2}{\sqrt{-g}}\frac{\delta S_{\mathrm{M}}}{\delta K^{\lambda }_{\;\;\mu\nu}},\label{xi3}
\eeq
but now with the connection being fully general. Using the identity 
\begin{equation}
    \Psi_{\lambda\mu\nu} \equiv \PP{_{\lambda\mu\nu}}{2}-\PP{_{\lambda\mu\nu}}{1},
\end{equation}
we can easily cast Eq.~\eqref{eq:ConnectionFE} into 
\begin{gather}
    \PP{_{\lambda\mu\nu}}{2} = 2\kappa \Xi_{\lambda\mu\nu} , \\
    \PP{_{\lambda\mu\nu}}{1} =- 2\kappa \Xi_{\lambda\mu\nu} 
\end{gather}
The composite tensor $\Psi_{\lambda}{}^{\mu\nu}$ was defined in~\eqref{psi}, and $\Xi_\lambda{}^{\mu\nu}$ in~\eqref{eq:XiDef}. Notice that since $\PP{_\lambda{}^{\lambda\nu}}{i}=0$, the field equations dictate that $\Xi_\lambda{}^{\lambda\mu}=0$ as already mentioned. Therefore, we may as well start with hypermomentum tensors already fulfilling this trace property.

The $i$-th distortion tensor can always be written as \cite{Iosifidis:2018jwu}
\begin{eqnarray}
    \NN{_{\lambda\mu\nu}}{i}&=&\frac12 \left( \PP{_{\lambda\mu\nu}}{i}-\PP{_{\mu\nu\lambda}}{i}-\PP{_{\nu\lambda\mu}}{i}\right)+\frac{1}{2(n-2)} g_{\mu\nu}\left( \PP{^{\alpha}{}_{\lambda\alpha}}{i} - \PP{_{\lambda\alpha}{}^{\alpha}}{i} \right)-\nonumber\\
    &&-\frac{1}{2(n-2)} g_{\lambda\nu}\left( \PP{^{\alpha}{}_{\mu\alpha}}{i} - \PP{_{\mu\alpha}{}^{\alpha}}{i} \right)+\frac12 g_{\lambda\mu}q_\nu{}^{(i)}.
\end{eqnarray}
Therefore, the solution to the field equations for the $i$-th connection is 
\begin{equation}
    \Gam{^\lambda{}_{\mu\nu}}{i}=\Tilde{\Gamma}^\lambda{}_{\mu\nu}+\kappa(-1)^iX^\lambda{}_{\mu\nu}{}+\frac12 \delta_{\mu}^\lambda q_\nu{}^{(i)},
\end{equation}
where
\begin{eqnarray}
    X_{\lambda\mu\nu} &:=&\XX{_{\lambda\mu\nu}}{i}-\XX{_{\mu\nu\lambda}}{i}-\XX{_{\nu\lambda\mu}}{i}+\frac{1}{n-2} g_{\mu\nu}\left( \XX{^{\alpha}{}_{\lambda\alpha}}{i} - \XX{_{\lambda\alpha}{}^{\alpha}}{i} \right)-\frac{1}{n-2} g_{\lambda\nu}\left( \XX{^{\alpha}{}_{\mu\alpha}}{i} - \XX{_{\mu\alpha}{}^{\alpha}}{i} \right).
\end{eqnarray}
It follows that the mean connection $\Gamma^\lambda{}_{\mu\nu}$ is 
\begin{equation}
    \Gamma^\lambda{}_{\mu\nu} \equiv \frac{1}{2}\left(\Gam{^\lambda{}_{\mu\nu}}{1}+\Gam{^\lambda{}_{\mu\nu}}{2}\right) = \Tilde{\Gamma}^\lambda{}_{\mu\nu}+\frac14 \delta^\lambda_\mu \left(q_\nu{}^{(1)}+q_\nu{}^{(2)}\right).
\end{equation}
Partially consuming the gauge freedom by setting 
\begin{equation}
    \xi_\mu{}^{(2)} = -\xi_\mu{}^{(1)}-\frac{1}{2}\left(q_\nu{}^{(1)}+q_\nu{}^{(2)}\right),
\end{equation}
we conclude that the mean connection is always the Levi-Civita connection (up to the choice of gauge). We can further set $\xi_\mu{}^{(1)}=-q_\nu{}^{(1)}/2$, for which gauge the solution acquires the final form
\begin{equation}
    \Gam{^\lambda{}_{\mu\nu}}{i}=\Tilde{\Gamma}^\lambda{}_{\mu\nu}+\kappa(-1)^i X^\lambda{}_{\mu\nu}{}\equiv \Tilde{\Gamma}^\lambda{}_{\mu\nu}+N^{\lambda \;\;\;(i)}_{\;\;\mu\nu}. \label{NNN}
\end{equation}
Note that the two connections annihilate each other, viz.,
\beq
N_{\alpha\mu\nu}^{\;\;\;\;(1)}+N_{\alpha\mu\nu}^{\;\;\;\;(2)}=0,
\eeq
and this constitutes a generalized geometry of which the statistical and torsional-statistical manifolds are certain subclasses.

Moreover, the metric field equations read
\begin{equation}
    \mathcal{R}_{(\mu\nu)}-\frac12 g_{\mu\nu}\mathcal{R} = \kappa T_{\mu\nu}.
\end{equation}
Performing a post-Riemannian expansion, we get 
\begin{eqnarray}
    \tilde{R}_{\mu\nu} -\frac12 g_{\mu\nu}\tilde{R} &=& \kappa T_{\mu\nu}+ \kappa^2\left( X^\alpha{}_{(\mu\nu)}X^\beta{}_{\alpha\beta} - X^{\beta}{}_{\alpha(\mu}X^\alpha{}_{\nu)\beta}\right)+\nonumber\\
    &&+\frac{\kappa^2}{2}g_{\mu\nu}\left( X^{\alpha\beta\gamma} X_{\beta\gamma\alpha} - X^{\alpha}{}_{\beta\alpha} X^{\beta\gamma}{}_\gamma\right),
\end{eqnarray}
whose Ricci form reads 
\begin{eqnarray}
    \tilde{R}_{\mu\nu} = \kappa \left( T_{\mu\nu}-\frac{1}{n-2} g_{\mu\nu} T\right)+\kappa^2\left( X^\alpha{}_{(\mu\nu)}X^\beta{}_{\alpha\beta} - X^{\beta}{}_{\alpha(\mu}X^\alpha{}_{\nu)\beta}\right).
\end{eqnarray}

Let us now assume that 
\begin{equation}
    \Xi_{\lambda\mu\nu} = \alpha C_{\lambda\mu\nu}+\beta A_{\lambda\mu\nu},
\end{equation}
with $C_{\lambda\mu\nu}$ being a completely symmetric and trace-free tensor, $A_{\lambda\mu\nu}$ being a 3-form field, and $\alpha,\beta$ real numbers. When this is the case, we have that
\begin{equation}
    X_{\lambda\mu\nu} = -\alpha C_{\lambda\mu\nu}-\beta A_{\lambda\mu\nu},
\end{equation}
and it follows that the connection solution is 
\begin{equation}
    \Gam{^\lambda{}_{\mu\nu}}{i}=\Tilde{\Gamma}^\lambda{}_{\mu\nu}+\kappa(-1)^{i+1} \left( \alpha C^\lambda{}_{\mu\nu}+\beta A^{\lambda}{}_{\mu\nu}\right).
\end{equation}
The $i$-th torsion reads 
\begin{equation}
    S_{\mu\nu\lambda}{}^{(i)} = (-1)^{i+1}\kappa\beta A_{\lambda\mu\nu},
\end{equation}
whereas the $i$-th nonmetricity is 
\begin{equation}
    Q_{\lambda\mu\nu}{}^{(i)} = 2(-1)^{i+1}\kappa \alpha C_{\lambda\mu\nu}.
\end{equation}
The metric field equations become
\begin{eqnarray}
    \tilde{R}_{\mu\nu} -\frac12 g_{\mu\nu}\tilde{R} &=& \kappa T_{\mu\nu}+(\alpha \kappa)^2\left( \frac{1}{2}g_{\mu\nu}C^2 -  C_\mu{}^{\alpha\beta}C_{\nu\alpha\beta}\right)+(\beta\kappa)^2\left(\frac12 g_{\mu\nu}A^2 - A_\mu{}^{\alpha\beta} A_{\nu\alpha\beta}\right),
\end{eqnarray}
with $C^2 = C_{\lambda\mu\nu}C^{\lambda\mu\nu}$ and $A^2 = A_{\lambda\mu\nu}A^{\lambda\mu\nu}$. Clearly, the previous models are recovered if we set $\alpha=0,\beta=1$ (TSM) or $\alpha=1,\beta=0$ (statistical manifold). Hence, we see that the general biconnection theory can accommodate both scenarios if we feed it with appropriate hypermomentum sources.

\section{Conclusions}

We have formulated a biconnection theory of gravity and have shown its correspondence with the geometry of statistical manifolds and torsional statistical manifolds, under certain assumptions. In particular, we started with a gravitational action given by the so-called mutual scalar curvature, a scalar depending on the two connections, constructed with the mutual curvature tensor recently defined in \cite{iosifidis2023torsion}. We then showed for the symmetric (i.e. vanishing torsion)  case that the vacuum theory is equivalent to GR, with the two connections coinciding with the Levi-Civita connection. 

Things become more interesting when matter is added. In order to solve the problem of which connection couples to matter and why, and also to have both connections on an equal footing, we consider couplings entering only through the difference tensor (see Eq. \ref{eq:diften}) of the two connections . For such couplings and for the symmetric case, we explicitly showed that if the primary hypermomentum (\ref{xi3}) is totally symmetric and trace-free, then the underlying geometry of the theory is that of a statistical manifold. Namely, there exists a totally symmetric tensor such that the two connections have the expressions (\ref{con1}) and (\ref{con2}), respectively. This totally symmetric tensor, called the cubic tensor in information geometry, is in our case the hypermomentum of matter. This intriguing correspondence could have some quite interesting applications since it would enable one to extract information about the microproperties of matter (hypermomentum) by studying statistical models, and vice versa\footnote{There is however one caveat here. The Fisher metrics derived from probability distributions (or more generally from divergencies) are positive definite in contrast to the Lorentzian metrics that appear in Gravity. One way to circumvent this problem would be to Wick-rotate one of the coordinates on the probability distributions but then one would face the unphysical issue of having complex valued probabilities. This problem is certainly quite interesting and is left for future work. }.

Furthermore, switching the roles of torsion and nonmetricity, we studied the metric version (i.e. vanishing nonmetricity) of the theory consisting of the mutual curvature and the aforementioned connection couplings. In this instance and given that the hypermomentum tensor is totally antisymmetric (i.e. fermionic matter), we explicitly showed that the underlying geometry is that of a torsional statistical manifold.

Finally, we formulated the general biconnection MAG, by allowing for two general connections that are neither metric, nor symmetric. Again, considering a gravitational sector consisting of the double contraction of the mutual curvature, we solved the connection field equations and arrived at a generalized geometry. More precisely, in this geometry the two connections have distortion tensors that differ by a sign and can accommodate both the statistical and torsional-statistical Manifolds for certain forms of the hypermomentum. It would be interesting to see if these geometric structures appear also in a dynamic way by enlarging the gravitational action (which in our case was Einstein-Hilbert-inspired) with more invariants built out of the mutual curvature and difference tensor. 

It would be interesting to explore the possibility of the emergence of this statistical structure from gravity in a dynamical way, namely, to find such a correspondence in vacuum by enlarging the gravitational sector of the biconnection theory. Additional invariants could be added to the action that are built out of the mutual curvature and the difference tensor.\footnote{Of course one could also add invariants constructed out of the curvature and/or torsion and nonmetricity tensors of the individual connections, however the presence of such terms would then spoil the symmetry of the gravitational action under the exchange of the two connections not placing them on equal footing anymore.}

\section{Acknowledgements}

D. I.'s work was supported by the Estonian Research Council grant (SJD14). K. P. acknowledges
financial support provided by the European Regional Development Fund (ERDF) through
the Center of Excellence TK133 “The Dark Side of the Universe” and PRG356 “Gauge
gravity: unification, extensions and phenomenology”. K.P. also acknowledges participation in the COST Association Action CA18108 “Quantum Gravity Phenomenology in the
Multimessenger Approach (QG-MM)”.

\appendix

\section{Variations\label{AppA}}

Let us gather here some useful variations that we used in order to derive the field equations of the biconnection Gravity. In the following both connections are fully  general. 

Firstly, for the variations of the curvature tensors of the two connections we have
	\beq
	T_{\lambda}^{\;\;\alpha\beta\gamma}\delta_{\Gamma^{(i)}}R^{\lambda(i)}_{\;\;\alpha\beta\gamma}=2\Big( \hat{\nabla}^{(i)}_{\alpha}T_{\lambda}^{\;\;\mu[\alpha\nu]}-T_{\lambda}^{\;\;\mu\alpha\beta}S_{\alpha\beta}^{\;\;\;\;\nu(i)} \Big)(\delta \Gamma^{\lambda(i)}_{\;\;\;\mu\nu}) 
	\eeq
	where   $T_{\lambda}^{\;\;\alpha\beta\gamma}$ is an arbitrary tensor field (or tensor density)   and  $\hat{\nabla}_{\alpha}^{(i)}=2S_{\alpha}^{(i)}-\nabla_{\alpha}^{(i)}$. Of course $\delta_{\Gamma^{(i)}}R^{\lambda(j)}_{\;\;\alpha\beta\gamma}=0$ for $i \neq j$. Continuing,
 for the term $-K^{\alpha}_{\;\;\rho[\gamma}K^{\rho}_{\;\;\;|\beta|\delta]}$ appearing in the mutual curvature we find
\beq
T_{\alpha}^{\;\;\beta\gamma\delta}\delta_{\Gamma^{(i)}}(-K^{\alpha}_{\;\;\rho[\gamma}K^{\rho}_{\;\;\;|\beta|\delta]})=(-1)^{i}\Big[ T_{\lambda}^{\;\;\beta[\nu\alpha]}K^{\mu}_{\;\;\beta\alpha}+T_{\alpha}^{\;\;\mu[\beta\nu]}K^{\alpha}_{\;\;\lambda\beta} \Big](\delta \Gamma^{\lambda (i)}_{\;\;\mu\nu})
\eeq
Such that the total variation of the mutual curvature with respect to the i-th connection reads (up to total derivatives)
\beq
T_{\alpha}^{\;\;\beta\gamma\delta}\delta_{\Gamma^{(i)}}\mathcal{R}^{\lambda}_{\;\;\alpha\beta\gamma}=\Big[ \hat{\nabla}^{(i)}_{\alpha}T_{\lambda}^{\;\;\mu[\alpha\nu]}-T_{\lambda}^{\;\;\mu\alpha\beta}S_{\alpha\beta}^{\;\;\;\;\nu(i)}+(-1)^{i}(T_{\lambda}^{\;\;\beta[\nu\alpha]}K^{\mu}_{\;\;\beta\alpha}+T_{\alpha}^{\;\;\mu[\beta\nu]}K^{\alpha}_{\;\;\lambda\beta})\Big](\delta \Gamma^{\lambda(i)}_{\;\;\;\mu\nu}) 
\eeq
For $T_{\lambda}^{\;\;\alpha\beta\gamma}=\delta_{\lambda}^{[\beta}g^{\gamma]\alpha}$ we get 
\beq
\delta_{\Gamma^{(i)}}\mathcal{R}=\frac{1}{2}\Big( P_{\lambda}^{\;\;\mu\nu (i)}+(-1)^{i}(\delta_{\lambda}^{\nu}K^{\mu\alpha}_{\;\;\;\alpha}+K_{\alpha\lambda}^{\;\;\;\alpha}g^{\mu\nu}-K^{\mu\nu}_{\;\;\;\lambda}-K^{\nu\;\;\;\mu}_{\;\;\lambda})\Big)\delta \Gamma^{\lambda(i)}_{\;\;\mu\nu}
\eeq
which is used in deriving the connection field equations.

\section{Parallel transport with two connections \label{sec:PT}}

First, we note that the effects of parallel-transporting vectors (twisting angle and breaking of attempted quadrilaterals) can be captured by using the action of commutators of covariant derivatives. Let us define the operator 
\begin{equation}
    P^{ij}_{\mu\nu} := \nabla_\nu{}^{(j)}\nabla_{\mu}{}^{(i)},
\end{equation}
whose action on a vector field $V^\mu$ is associated with the following parallel-transport picture:
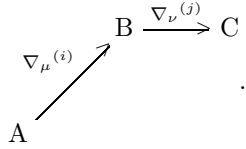
\begin{equation*}
\begin{diagram}
    & & \mathrm{B} & \rTo^{\nabla_\nu{}^{(j)}} &\mathrm{C}\\
     & \ruTo^{\nabla_\mu{}^{(i)}}  \\
    \mathrm{A}
\end{diagram}.
\end{equation*}
Here, the vector is parallel-transported from point A to B using the $i$-th connection with the resulting vector being further parallel-transported to a point C using the $j$--th connection.

At the same time, $P_{\nu\mu}^{ji}V^\lambda$ is associated with the following picture: 
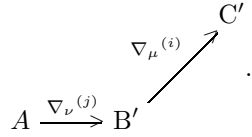
\[
\begin{diagram}
    &&&&\mathrm{C}'\\
&&&\ruTo^{\nabla_\mu{}^{(i)}}\\
A&\rTo^{\nabla_\nu{}^{(j)}}&\mathrm{B}'
\end{diagram}.
\]
If both connections are flat, torsion-free, and compatible with the metric, i.e., $\nabla_\mu{}^{(i)}=\partial_\mu$, then $(P^{ji}_{\nu\mu}-P^{ij}_{\mu\nu})V^\lambda=0$. For general connections on the other hand, we have 
\begin{equation}
    2P^{ij}_{[\nu\mu]}V^\lambda = R^\lambda{}_{\alpha\mu\nu}{}^{(i)}V^\alpha +2 S_{\mu\nu}{}^{\alpha(j)}\nabla_\alpha{}^{(i)}V^\lambda-2(-1)^jK^\lambda{}_{\alpha[\mu}\nabla_{\nu]}{}^{(i)}V^\alpha,
\end{equation}
which for $i=j$ assumes the usual form 
\begin{equation}
    2P^{ii}_{[\nu\mu]}V^\lambda = R^\lambda{}_{\alpha\mu\nu}{}^{(i)} V^\alpha + 2 S_{\mu\nu}{}^{\alpha(i)}\nabla_\alpha{}^{(i)}V^\lambda,
\end{equation}
due to the vanishing of the difference tensor. 
    
Now, observe that 
\begin{equation}
    \frac{1}{2}\sum_{i,j} P^{ij}_{[\nu\mu]}V^\lambda = R^\lambda{}_{\alpha\mu\nu}V^\alpha+\left( S_{\mu\nu}{}^{\alpha(1)}+S_{\mu\nu}{}^{\alpha(2)}\right)\nabla_\alpha V^\lambda,\label{eq:PSumOverAll}
\end{equation}
where $\nabla$ and $R^\lambda{}_{\rho\mu\nu}$ are given by eqs.~\eqref{eq:conaddition} and~\eqref{eq:CurvatureConvex}, respectively, for $t=1/2$, i.e., they stand for the mean connection and its curvature. The above schematically corresponds to summing over all possible ways to do the path AB'C' and subtracting it from the summation over all possible ways to do ABC; this, of course, divided by the number of possible ways to do each path which is four in the case of two connections. This complicated interpretation is utterly equivalent to attempting a closed loop via $\nabla$-parallel translations. Clearly, for connections with $S^{(1)}=-S^{(2)}$, namely vanishing mean torsion, quadrilaterals are preserved although both connections have torsion. If we further take our mean connection to be flat, then the right hand side of Eq.~\eqref{eq:PSumOverAll} vanishes completely, a result tantamount to parallel-transporting $V^\mu$ in flat spacetime, although here both connections have curvature and torsion. 

However, do also observe that 
\begin{equation}
    \sum_{i,j} P^{ij}_{[\nu\mu]}V^\lambda = \sum_{i\neq j} P^{ij}_{[\nu\mu]}V^\lambda+\sum_{i} P^{ii}_{[\nu\mu]}V^\lambda,
\end{equation}
where 
\begin{subequations}\label{eq:PSumOverDifferent}
\begin{eqnarray}
    \sum_{i\neq j} P^{ij}_{[\nu\mu]}V^\lambda &=& \mathcal{R}^\lambda{}_{\alpha\mu\nu}V^\alpha+\left(S_{\mu\nu}{}^{\alpha(2)}\nabla_\alpha{}^{(1)}+ S_{\mu\nu}{}^{\alpha(1)}\nabla_\alpha{}^{(2)} \right)V^\lambda,\label{eq:ineqj}\\
    \sum_{i} P^{ii}_{[\nu\mu]}V^\lambda &=& \frac{1}{2}\left(R^\lambda{}_{\alpha\mu\nu}{}^{(1)}+R^\lambda{}_{\alpha\mu\nu}{}^{(2)}\right)V^\alpha+\left(S_{\mu\nu}{}^{\alpha(1)}\nabla_\alpha{}^{(1)}+ S_{\mu\nu}{}^{\alpha(2)}\nabla_\alpha{}^{(2)} \right)V^\lambda,
\end{eqnarray}
\end{subequations}
with $\mathcal{R}^\lambda{}_{\rho\mu\nu}$ being the mutual curvature tensor given in~\eqref{eq:mutualcurvature}. Equation~\eqref{eq:ineqj} schematically amounts to summing over all possible ways to do the path AB'C' using different connections and subtracting it from the summation over all possible ways to do ABC, again, using different connections; this divided by the number of possible ways to do each path in this fashion which is of course equal to the number of connections. Further manipulating~\eqref{eq:PSumOverDifferent}, we can show that 
\begin{subequations}\label{eq:PSumOverDifferentAlt}
\begin{eqnarray}
    \sum_{i\neq j} P^{ij}_{[\nu\mu]}V^\lambda &=& \left(\mathcal{R}^\lambda{}_{\alpha\mu\nu} -\frac{1}{2} K^\rho{}_{[\mu\nu]}K^\lambda{}_{\alpha\rho}\right) V^\alpha + \left(S_{\mu\nu}{}^{\alpha(1)}+S_{\mu\nu}{}^{\alpha(2)}\right)\nabla_\alpha V^\lambda,\label{eq:ineqjAlt}\\
    \sum_{i} P^{ii}_{[\nu\mu]}V^\lambda &=& \frac{1}{2}\left(R^\lambda{}_{\alpha\mu\nu}{}^{(1)}+R^\lambda{}_{\alpha\mu\nu}{}^{(2)} + K^\rho{}_{[\mu\nu]}K^\lambda{}_{\alpha\rho}\right) V^\alpha+ \left(S_{\mu\nu}{}^{\alpha(1)}+S_{\mu\nu}{}^{\alpha(2)}\right)\nabla_\alpha V^\lambda,
\end{eqnarray}
\end{subequations}
where $K^\lambda{}_{[\mu\nu]} = S_{\mu\nu}{}^{\alpha(1)} - S_{\mu\nu}{}^{\alpha(2)}$, and we recall that 
\begin{equation}
    \mathcal R^\lambda{}_{\rho\mu\nu}= R^\lambda{}_{\rho\mu\nu}-\frac{1}{2}K^\lambda{}_{\sigma[\mu}K^{\sigma}{}_{|\rho|\nu}.
\end{equation}
Observe that when we split the overall sum into the above distinct pieces, the otherwise consistent identifications of curvature with the tensor contracted with $V$ and torsion with the tensor contracted with a covariant derivative of $V$ do not make sense for each piece separately. 

For example, looking at expression~\eqref{eq:ineqjAlt}, we can attribute the failure to close quadrilaterals to the torsion of the mean connection. However, the presence of a nontrivial twisting angle cannot be ascribed to a curvature of some connection, for the quantity contracted with $V$ is definitely not such a construct. Hence, in the case of attempting to close a loop with sequential $\nabla^{(i)}$-parallel and $\nabla^{(j)}$-parallel translations for $i\neq j$, we see that, if-f the two connections are torsion-free, then the tensor sourcing the nontrivial twisting angle is $\mathcal{R}^\lambda{}_{\rho\mu\nu}$. This justifies the name mutual ``curvature''. Nevertheless, in the presence of torsion, things turn out to be slightly ambiguous. To see this, assume for example that $S^{(1)}=-S^{(2)}$ and $\mathcal{R}^\lambda{}_{\rho\mu\nu}=0$. Using expression~\eqref{eq:ineqj}, it seems that the vector is not ``rotated'' and that infinitesimal quadrilaterals are not preserved. At the same time, looking at expression~\eqref{eq:ineqjAlt}, we stumble upon the opposite picture, namely that infinitesimal quadrilaterals are preserved, but the twisting angle is nontrivial, sourced by the nonvanishing torsion of the connections.

\bibliographystyle{unsrt}
\bibliography{refs}

\end{document}